\documentclass[runningheads]{llncs}

\usepackage[T1]{fontenc}
\usepackage{cite}
\usepackage{amsmath,amssymb,amsfonts}
\usepackage{algorithmic}
\usepackage{graphicx}
\usepackage{textcomp}
\usepackage[dvipsnames]{xcolor}
\usepackage{booktabs}
\usepackage{pifont}
\usepackage[most]{tcolorbox}
\usepackage[colorlinks=true,linkcolor=blue,citecolor=blue,urlcolor=blue]{hyperref}
\usepackage[frozencache,cachedir=.]{minted}
\usepackage{listings}
\usepackage{multirow}
\usepackage{placeins}
\usepackage{orcidlink}
\usepackage{subcaption}
\usepackage{fancyhdr}

\definecolor{pmdcolor}{HTML}{FFE6CC}
\definecolor{crccolor}{HTML}{F4F5FC}
\definecolor{crccolordark}{HTML}{91BEFF}

\tcbset{
    pmd/.style={
      on line,
      boxrule=0pt, 
      boxsep=0pt, 
      top=2pt,
      left=2pt, 
      bottom=2pt,
      right=1pt,
      colback=pmdcolor,
      colframe=black,
      coltext=blue!10!black,
      fontupper=\small\ttfamily
    },
    crc/.style={
      on line,
      boxrule=0pt, 
      boxsep=0pt, 
      top=2pt,
      left=2pt, 
      bottom=2pt,
      right=1pt,
      colback=crccolor,
      colframe=black,
      coltext=blue!80!black,
      fontupper=\small\ttfamily
    }
}

\newtcolorbox{titlebluebox}[2][]{
  colback=crccolordark!5!white,
  colframe=crccolordark!75!black,
  title=\textbf{#2},
  #1
}

\newcommand{\boldparagraph}[1]{\vspace{1em} \noindent \textbf{#1}}

\begin{document}

\pagestyle{headings}
\fancyhf{}
\renewcommand{\headrulewidth}{0pt}
\setlength{\headheight}{26pt}
\setlength{\headsep}{15pt}
\fancyhead[C]{\textcolor{gray}{\footnotesize This is the author’s version of a paper accepted for publication
at the International Semantic Web Conference (ISWC) 2026. The final version will be published by Springer.}}

\title{Integrating Semantics into Research Data Management: Modelling and Validating\\ Materials Science Experiment Workflows}
\titlerunning{Modelling and Validating Materials Science Experiment Workflows}



\author{Samuel García Vázquez\inst{1}\orcidlink{0009-0006-0232-8677} 
\and Victor Dudarev\inst{2}\orcidlink{0000-0001-7243-9096} 
\and Alfred Ludwig\inst{2}\orcidlink{0000-0003-2802-6774} 
\and\\ Markus Stricker\inst{2}\orcidlink{0000-0002-8933-0238}
\and Maribel Acosta\inst{1}\orcidlink{0000-0002-1209-2868}
}

\authorrunning{García Vázquez et al.}

\institute{Technical University of Munich, Germany\\
\email{\{samuel.garcia,maribel.acosta\}@tum.de}\\
\and Ruhr University Bochum, Germany\\
\email{\{victor.dudarev,alfred.ludwig,markus.stricker\}@rub.de}}

\maketitle
\thispagestyle{fancy}

\begin{abstract}
The incorporation of Semantic Web technologies within scientific environments is becoming an increasingly popular Research Data Management (RDM) practice. While ontologies offer flexible, reusable and machine-readable vocabularies to describe domain-specific research data, Knowledge Graphs (KGs) facilitate the integration of heterogeneous data sources into an interoperable collection. Furthermore, KGs offer additional advantages, notably the use of  expressive SPARQL queries, or the ability to define complex data validation rules with SHACL. This work describes the modelling of a relational RDM system with an ontology, and the subsequent construction of a KG based on it. Enabled by the highly interconnected nature of research data and experiment workflows present in the system, we not only show how we can easily and reliably build an efficient KG from such domain-specific RDM systems, but also how doing so enables more advanced use cases. This is demonstrated by the modelling of \emph{ideal} counterparts for the experiment workflows logged in the KG, which are then used to programmatically generate SHACL shapes that fully validate the conformance of the latter. By integrating this functionality within a UI, we allow researchers to plan, reuse, share, and track the progress of their daily experiments. 

\end{abstract}



\section{Introduction}

The increasing digitalization of research environments in the past years has highlighted the need for careful storage and curation of all experimental data. To achieve this, the use of Research Data Management (RDM) practices is becoming necessary, and is usually facilitated by RDM systems (e.g., \cite{fitschenCaosDBResearchData2019a, bauchOpenBISFlexibleFramework2011, coscine}). 
These systems allow for capturing not only research data, but also provenance metadata, describing details such as its source, the procedure it has undertaken to be generated, or the software tools and equipment used to obtain it. Within the field of Materials Science, the need for RDM practices is exacerbated by the complexity of the data generated, whose highly heterogeneous nature can range from atomistic simulations to the characteristics of synthesized materials, or detailed records of all processing steps and experiments performed on them. As a result, specialized RDM systems have been developed to take into account the characteristics of data in this field, particularly MatInf~\cite{matinf} or NOMAD~\cite{nomad}. 
In addition, studying certain hypotheses in Materials Science requires executing complex experiment workflows that must record all processing steps and experiments performed, involving the coordination across different labs.

Up until now, most research work regarding the integration of Semantic Web technologies within RDM was performed to improve the reusability and interoperability of research data, and thus its compliance with the FAIR principles~\cite{wilkinsonFAIRGuidingPrinciples2016}. In the Materials Science and Chemistry domains alone, we can already find several works applying KGs for the integration of otherwise non-interoperable or isolated data sources~\cite{hernandezDataIntegrationFramework2024, bayerleinSemanticIntegrationDiverse2024, schillingSeamlessScienceLifting2025, vollbrechtIntegratedDataPipeline2025}. Moreover, state-level RDM initiatives have also started promoting them, as in the case of the \emph{Nationale Forschungsdateninfrastruktur} (NFDI)~\cite{rdmsolutionsnfdi} in Germany. Yet, the use of these technologies can also unveil advanced use cases that allow for exploiting research data in novel ways.

\boldparagraph{Problem Statement}
As part of our work in the Collaborative Research Center (CRC) 1625\footnote{\href{https://www.ruhr-uni-bochum.de/crc1625/}{Collaborative Research Center (CRC) 1625}.}, an RDM system, MatInf~\cite{matinf}, is currently deployed to facilitate and foster the collaboration and sharing of research data across its participating institutions. 
With a generic but powerful relational data schema, this system is carefully designed to accommodate different research scenarios in Materials Science. Nonetheless, this design poses a considerable challenge: the complexity of the data schema does not allow its end users to programmatically access data via structured queries, requiring the development of APIs and interfaces instead. Moreover, querying any data structures that naturally form graphs in relational databases, such as experiment workflows of any kind, requires complex queries that lead to performance degradation.
Consequently, establishing a \emph{semantic layer} over the underlying system would facilitate access to and interaction with the research data, while also enabling new use cases for graph structures that were previously too complex to leverage in relational databases~\cite{poster_paper, journal_paper}.

\boldparagraph{Our Solution}
In this work, we describe the process of designing an ontology and then constructing a Knowledge Graph (KG) that represents the research data stored in the MatInf RDM system, alongside novel use cases for managing workflows to model experiments and plans. Our key contributions are:
\begin{itemize}
    \item We prove the feasibility of modelling, building and maintaining a live semantic layer, by deploying a KG with a tailored ontology on top of an existing, domain-specific RDM system.

    \item We showcase additional use cases offered by the inherent graph nature of research data and experiment workflows. This is demonstrated by modelling rich and reusable experiment plans in the KG, which we can then use to programmatically track and validate the recorded experiment workflows and research data in the system.

    \item We demonstrate the benefits that this semantic layer brings to the table in RDM, offering an analysis of querying performance and usability of the resulting KG against the existing relational database.
\end{itemize}

The rest of this work is structured as follows. Section~\ref{sec:preliminaries} describes our original research data. Next, Section~\ref{sec:ontology_development} discusses the modelling of our ontology, while Section~\ref{sec:kgc} describes the KG construction process.  Section~\ref{sec:evaluation} discusses the evaluation and lessons learned, and   Section~\ref{sec:rel_work} describes the related work. 
Finally, Section~\ref{sec:conclusion_future_work} summarizes our conclusions and future research avenues.

\section{Preliminaries}
\label{sec:preliminaries}

\subsection{Research Data Description}
\label{sec:nature_of_the_data}
One peculiarity of our domain is the highly heterogeneous research data that is currently being generated and stored. Due to this, we must firstly understand the different categories of data before representing it within an ontology:

\boldparagraph{Research Objects}
The tasks in our domain comprise the simulation, synthesis, and experimental analysis of thin-film and Compositionally Complex Solid Solution (CCSS) surfaces. 
These commonly consist of five or more different elements mixed in a simple single-phase crystal structure, which hold promising electrocatalytic properties. In order to enable the high-throughput characterization of these materials, a \emph{materials library} is synthesized as a single composition-spread thin film, which holds a collection of several hundreds of material compositions and structures to be analyzed~\cite{ludwigDiscoveryNewMaterials2019}. A materials library can be further split into multiple \emph{samples}, or be transformed by, e.g., annealing them. Furthermore, an interesting experimental result on any of these two objects may lead to the synthesis of a similar materials library with, e.g., a different underlying substrate. Due to this, materials libraries and samples are highly interconnected. For simplicity, we will refer to either of them as \emph{research objects}.

\boldparagraph{Experiment Results}
An \emph{experiment result} comprises any kind of data related to a research object that has been recorded in MatInf. Currently, there are more than 20 different types of computational simulations and electrochemical, surface and volume composition characterization techniques being performed in our project. As a result, we require the use of parsers for binary file formats to extract metadata or to convert them to structured formats. This is facilitated by MatInf, which will then hold references to these files in its database records. Aside from these results, MatInf additionally allows uploading photos, simulation results, reports, publication references and other miscellaneous documents, which may be interconnected with other objects or results to provide context.

\boldparagraph{Measurement Areas and Compositions}
Although most characterization techniques only contain a reference to raw and/or parsed files in MatInf, one exception to this is the \emph{Energy-Dispersive X-ray spectroscopy} (EDX) technique. EDX yields the volume composition of research objects (i.e., the overall percentage of atoms of each element within it). This technique is applied on 342 different measurement areas within the materials library, or a subset of them in the case of samples. As a result, an EDX measurement uploaded to the system is actually composed of up to 342 \textit{sub-measurements}, whose data is then represented as chemical compositions linked to their corresponding measurement areas. This is planned to be applied to other measurements in the future, and as a result we need to facilitate the representation of \emph{measurement hierarchies} with increasingly higher levels of detail.

\boldparagraph{Experiment Workflows}
As there are more than ten different projects within the CRC 1625 synthesizing, analyzing or modifying research objects, tracking \emph{what}, \emph{when}, and \emph{by whom} anything was done to them (i.e., provenance metadata) is crucial. To achieve this, MatInf logs extensive \emph{handover} events. A handover represents the act of transferring a research object from one person to another, usually with the goal of performing some characterization technique or treatment on it. These transfers typically involve different research projects from both Electrochemistry and Materials Science and, thus, different locations (labs). By chaining the handovers of an object chronologically and linking them to experiment results, we can form an \emph{experiment workflow}. This, as a whole, forms an ordered and detailed record of all computational or physical activities performed on an object.  Note that, although a handover can only involve an individual materials library or a number of related samples, existing relations between the latter (e.g., between a materials library and its derived samples) can effectively connect these workflows.

\subsection{Original Data Source Structure}
\label{sec:db_schema}
All (meta-)data in MatInf, with the exception of raw or processed files, is currently stored in a relational database. 
This database schema is designed to be \emph{general-purpose} to accommodate any domain-specific scenario in Materials Science. 
This is achieved by treating anything uploaded or logged within MatInf as an \emph{object} within an \emph{ObjectInfo} table. These objects are assigned a \emph{type} (e.g., \emph{Sample}, \emph{EDX measurement} or \emph{Handover}), which are separately stored in an extensible \emph{TypeInfo} table. 
Moreover, arbitrary connections between two objects are allowed and stored in a \emph{ObjectLinkObject} table, enabling this way graph structures. Finally, it allows the creation of dictionary structures via labelled attributes to objects with string, integer and float property tables, and offers dedicated tables for compositions, publications, handovers, and research objects. 

This essentially means that we can coalesce the different tabular and dictionary structures into a directed, labelled graph\footnote{Edge labelling is only used in MatInf to, e.g., signal whether an edge was established automatically, or to indicate why two research objects are linked, lacking an ontology.} and, therefore, into a KG. This way, edges and object types will be modelled by the ontology, which will additionally offer richer class hierarchies for its types.

\section{The CCSS Ontology}
\label{sec:ontology_development}

As part of our work, we have designed the \emph{CCSS ontology} to model all the data structures designed to support CCSS (Compositionally Complex Solid Solutions) exploration in MatInf, and also the characterization technique vocabulary used for this means. Following the standard practices in ontology development, we have opted to reuse and extend existing ontologies where possible, in order to foster the interoperability of our data. 
First, we describe how we selected an ontology from the literature (\S\ref{sec:ontology_selection}). 
Next, we present the modelling of the main concepts from our domain (cf. Section~\ref{sec:preliminaries}):   
research objects and experiment results (\S\ref{sec:modeling_research_data}), experiment workflows (\S\ref{sec:modeling_handover_workflows}), 
and, in addition, experiment plans (\S\ref{sec:handover_workflow_models}).

\subsection{Ontology Selection}
\label{sec:ontology_selection}

Given the nature of our project, our main objectives and core requirements for developing an ontology are 1)~the reusability and interoperability with other Materials Science ontologies, and 2)~the modelling of MatInf-specific concepts. However, the landscape of ontologies in Materials Science is under active development, lacking a fully standardized and widely accepted general ontology for the field. Moreover, we need to represent highly domain-specific and heterogeneous data, including experiment workflows, characterization techniques, and provenance metadata. Therefore, we developed a set of requirements (Table~\ref{tab:ontology_cqs}), and chose and extended the best-fitting ontology -- including top-level (TLO), medium-level (MLO), and domain-specific ontologies (DSO) -- available in the literature.  

\begin{itemize}
    \item \textbf{Platform MaterialDigital core ontology (PMDco)}: Developed by the consortium of the same name\footnote{\href{https://www.materialdigital.de/}{Platform MaterialDigital} (Last accessed: July 2026).}, PMDco~\cite{PMDco} is an MLO that specializes in processes, experiments, and computational workflows in Materials Science. It extends the established PROV-O TLO~\cite{PROVO} to model provenance metadata, and relies on the DSOs ChEBI~\cite{chebi} and QUDT~\cite{FAIRsharing_QUDT_2025} to model chemical elements  and measurement units, respectively.

    \item \textbf{Elementary Multiperspective Material Ontology (EMMO)}: Developed by the European Materials Modelling Council\footnote{\href{https://emmc.eu/}{European Materials Modelling Council} (Last accessed: July 2026).}, EMMO~\cite{EMMO} is a TLO with a focus on materials modelling and characterization techniques. It also counts with a wide collection of MLOs and DSOs that expand on its concepts, such as CHAMEO~\cite{delnostroCHAMEOOntologyHarmonisation2022} for the materials characterization domain.

    \item \textbf{BWMD ontology}: Similarly to PMDco, BWMD~\cite{bwmd} was developed within Platform MaterialDigital, and is focused on process chains and material structure descriptions. In this case, it extends BFO~\cite{bfo}.

    \item \textbf{Materials Science and Engineering Ontology (MSEO)}: Developed by the Materials Open Laboratory platform\footnote{\href{https://mat-o-lab.github.io/OrgSite/}{Materials Open Laboratory} (Last accessed: July 2026).}, MSEO\footnote{\href{https://github.com/Mat-O-Lab/MSEO}{Materials Science and Engineering Ontology (MSEO)} (Last accessed: July 2026)} is an MLO focused on representing experimental data. It extends the IOF~\cite{IOF} ontology.
\end{itemize}

\begin{table}[t]
\centering
\tiny
\caption{Requirements and conformance of the best candidates for our ontology to be based on. Green ticks indicate conformance, yellow tildes indicate partial conformance, and red crosses indicate non-conformance.}
\begin{center}
\begin{tabular}{ll@{\hskip 2em}cccc}
\toprule
\textbf{ID} & \textbf{Requirement}                                                           & \textbf{PMDco} & \textbf{EMMO} & \textbf{BWMD} & \textbf{MSEO} \\ \midrule
R1         & The ontology must be able to express users and groups                                  & \textbf{\color{ForestGreen}\ding{51}}              & \textbf{\color{ForestGreen}\ding{51}}             & \textbf{\color{ForestGreen}\ding{51}}             & \textbf{\color{ForestGreen}\ding{51}}                  \\ \midrule
R2         & The ontology must be able to express experimental results         & \textbf{\color{ForestGreen}\ding{51}}              & \textbf{\color{ForestGreen}\ding{51}}             & \textbf{\color{ForestGreen}\ding{51}}             & \textbf{\color{ForestGreen}\ding{51}}                  \\ \midrule
R3         & The ontology must be able to express experiment workflows              & \textbf{\color{ForestGreen}\ding{51}}              & \textbf{\color{ForestGreen}\ding{51}}             & 
$\textcolor{orange}{\boldsymbol{\sim}}$            & $\textcolor{orange}{\boldsymbol{\sim}}$                  \\ \midrule
R4         & The ontology interconnects elements from CQ1-CQ3                   & \textbf{\color{ForestGreen}\ding{51}}              & \textbf{\color{ForestGreen}\ding{51}}             & \textbf{\color{ForestGreen}\ding{51}}             & \textbf{\color{ForestGreen}\ding{51}}                  \\ \midrule
R5         & The ontology models provenance metadata  & \textbf{\color{ForestGreen}\ding{51}}              & \textbf{\color{ForestGreen}\ding{51}}             & $\textcolor{orange}{\boldsymbol{\sim}}$        & \textbf{\color{ForestGreen}\ding{51}}                  \\ \midrule
R6         & The ontology is aligned with the field of Materials Science                           & \textbf{\color{ForestGreen}\ding{51}}              & \textbf{\color{ForestGreen}\ding{51}}             & \textbf{\color{ForestGreen}\ding{51}}             & \textbf{\color{ForestGreen}\ding{51}}                  \\ \midrule
R7         & The ontology must extend other existing, widely used ontologies                              & \textbf{\color{ForestGreen}\ding{51}} (PROV-O)  & \textbf{\color{BrickRed}\ding{55}} (TLO)             & \textbf{\color{ForestGreen}\ding{51}} (BFO)           & \textbf{\color{ForestGreen}\ding{51}} (IOF)             \\ \midrule
R8         & The ontology is finalized or, if not, offers recent stable releases         & \textbf{\color{ForestGreen}\ding{51}}              & \textbf{\color{ForestGreen}\ding{51}}             & \textbf{\color{ForestGreen}\ding{51}}             & \textbf{\color{BrickRed}\ding{55}}                  \\ \midrule
R9         & The ontology is currently used by a wide community                   & \textbf{\color{ForestGreen}\ding{51}}              & \textbf{\color{ForestGreen}\ding{51}}             & \textbf{\color{ForestGreen}\ding{51}}             & $\textcolor{orange}{\boldsymbol{\sim}}$                 \\ \bottomrule
\end{tabular}
\end{center}
\label{tab:ontology_cqs}
\end{table}

As shown in Table~\ref{tab:ontology_cqs}, PMDco, at its version 2.0.7, met all established criteria, and is the ontology we chose to extend. This choice is additionally based on existing comparisons from literature: As shown by Norouzi et al.~\cite{norouziLandscapeOntologiesMaterials2024a} and Beygi Nasrabadi et al.~\cite{ebrahimperformanceeval}, PMDco proves to be a very balanced ontology in terms of design, semantic richness and query complexity.

Throughout the examples, our ontology will be prefixed as \tcbox[crc]{crc:ClassName}, corresponding to \href{https://crc1625.mdi.ruhr-uni-bochum.de/}{https://crc1625.mdi.ruhr-uni-bochum.de/}. Similarly, PMDco will be prefixed as \tcbox[pmd]{pmdco:className}, corresponding to \href{https://w3id.org/pmd/co}{https://w3id.org/pmd/co}.

\subsection{Modelling Research Objects and Experiment Results}
\label{sec:modeling_research_data}
Research objects are implemented as subclasses of \tcbox[pmd]{pmdco:Object}, while experiment results are represented as subclasses of \tcbox[pmd]{pmdco:ValueObject}. A research object containing the more complex EDX \emph{sub-measurements} is \tcbox[pmd]{pmdco:composedOf} \tcbox[crc]{crc:MeasurementArea} instances, while the individual composition data is represented as \tcbox[pmd]{pmdco:ChemicalComposition} instances, each attached to the respective area it belongs to, and comprised of one \tcbox[pmd]{pmdco:ChemicalObject} per chemical element. This design aims to facilitate querying and avoid data duplication, as there may be multiple composition measurements taken on the same \tcbox[crc]{crc:MeasurementArea} instance.

\begin{figure}[t]
    \centering
    \includegraphics[width=\textwidth]{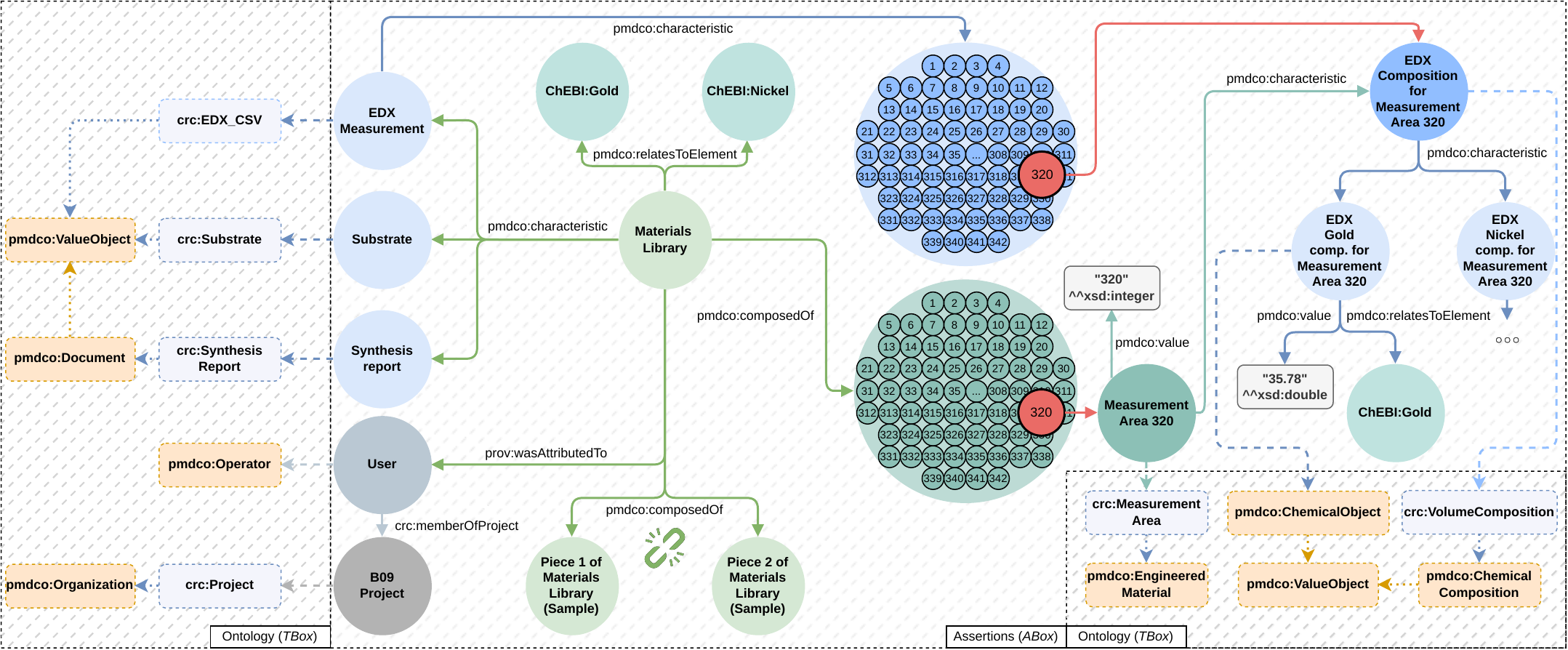}
    \caption{Overview of measurements and compositions of a materials library. Dashed and dotted arrows indicate class and subclass memberships, respectively.}
    \label{fig:sample_measurements}
\end{figure}

Entities in any of these categories hold provenance records (e.g., user attribution, creation dates, or MatInf identifiers), and, similarly to MatInf, hold references to files or raw values via the \tcbox[pmd]{pmdco:value} property. An example of research objects and experiment results can be found in Figure~\ref{fig:sample_measurements}. 

\subsection{Modelling Experiment Workflows}
\label{sec:modeling_handover_workflows}
\begin{figure}[t]
    \centering
    \includegraphics[width=\linewidth]{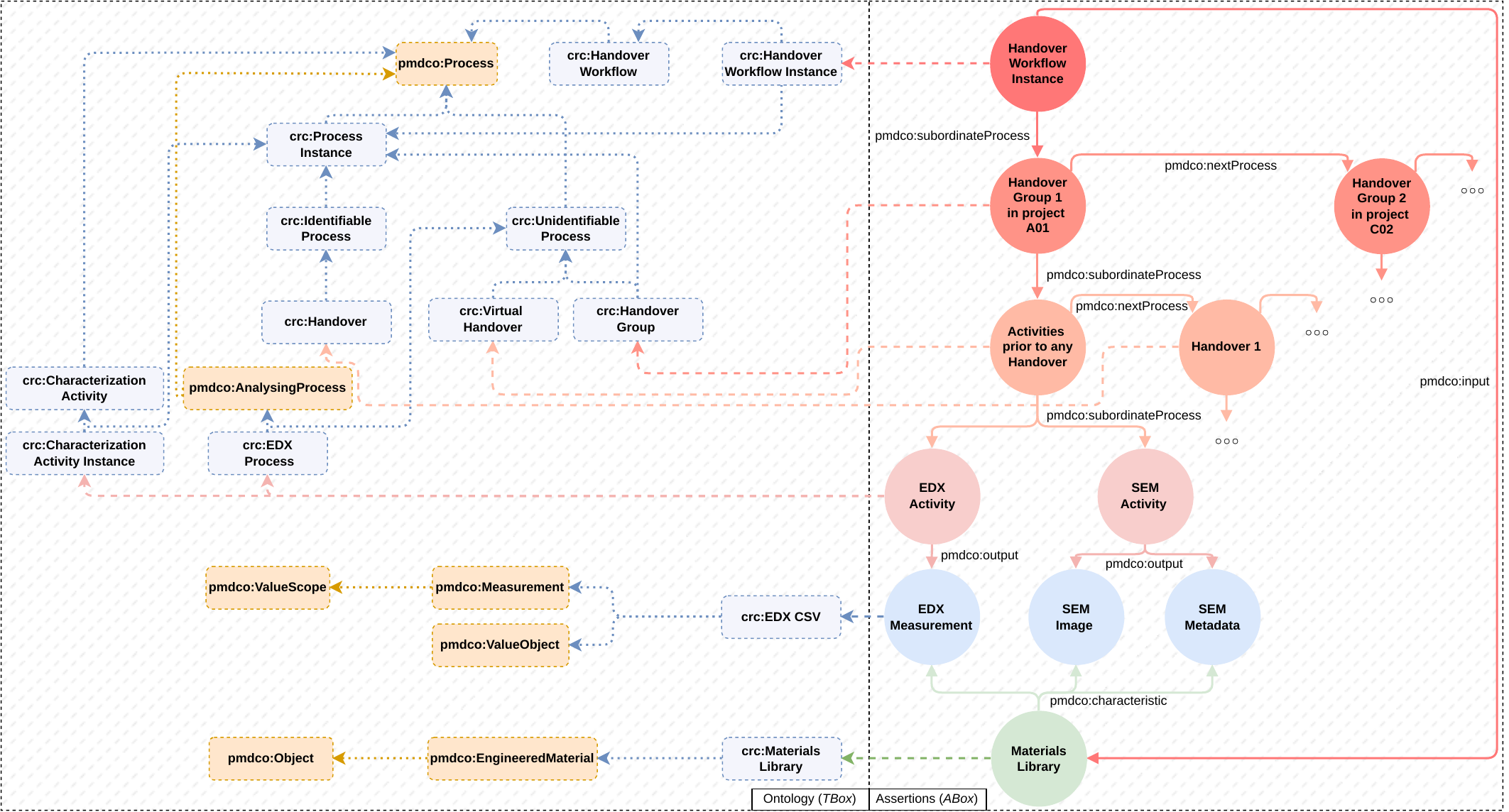}
    \caption{Overview of the experiment workflow for a materials library. Dashed and dotted arrows indicate class and subclass memberships, respectively.}
    \label{fig:workflow_instances}
\end{figure}

With the aim of effectively querying, visualizing, and, as will be described in Section~\ref{sec:handover_workflow_models}, validating these experiment workflows via SHACL, we have expanded the handover event chains present in MatInf to a 3-level hierarchy. As shown in Figure~\ref{fig:workflow_instances} and described below, a \tcbox[crc]{crc:HandoverWorkflowInstance} consists of the following elements, modelled as subclasses of \tcbox[pmd]{pmdco:Process}:

\begin{itemize}
    \item \textbf{Level 1}: The first level is formed by \tcbox[crc]{crc:HandoverGroup} entities, which contain a chain of $n \geq 1$ handovers that took place consecutively in time and within the same research project (i.e., the object was passed around members of the same project). Formally, given a handover $H_{n}$, directed at project $P$ and belonging to the handover group $HG$, if its originally succeeding handover $H_{n+1}$ was directed to a project $P' \neq P$, a new handover group $HG'$ is created for $H_{n+1}$. $HG'$ then contains all consecutive handovers starting from $H_{n+1}$ that were directed to $P'$, until the same process repeats. The original chronological ordering is preserved by establishing that $HG$ has as its \tcbox[pmd]{pmdco:nextProcess} $HG'$.

    \item \textbf{Level 2}: This level consists of \tcbox[crc]{crc:Handover} entities, representing the same concept as in MatInf. The first handover of each group, chronologically, will be a \tcbox[pmd]{pmdco:subordinateProcess} of its corresponding \tcbox[crc]{crc:HandoverGroup}. Likewise, consecutive handovers are connected to their immediate successors within the group via \tcbox[pmd]{pmdco:nextProcess}. Additionally, we model a special \tcbox[crc]{crc:VirtualHandover}, representing the work performed on the materials library or sample before any handover was logged in the system (i.e., preliminary data such as photos or reports). This will always be present as the initial handover of the first handover group.

    \item{} \textbf{Level 3}: The last level consists of \tcbox[crc]{crc:CharacterizationActivity} entities, representing the act of uploading any number of measurements belonging to the same measurement technique, research object, and time frame. Activities serve to simplify the visualization of measurements in this common occurrence, as the same measurement technique may involve the upload of multiple raw and processed files, each appearing as individual entities. More importantly, this relieves end users from using \texttt{a\textbackslash rdfs:subClassOf*} patterns in queries, were this is to be modelled instead as a class hierarchy. Formally, an activity assigned to $H_n$ will contain all such measurements that were uploaded between the time the handovers $H_n$ and $H_{n+1}$ (if it exists) took place. In order to indicate the characterization technique they belong to, they are also part of one of the custom subclasses of \tcbox[pmd]{pmdco:AnalysingProcess} (e.g., \tcbox[crc]{crc:EDXProcess}). 
\end{itemize}

This 3-level hierarchy allows us to query and visualize any of the elements in the experiment workflow in varying levels of detail, while maintaining a clear separation between experiment workflows and the actual measurement data: We can query for the measurements of an object based on its \tcbox[pmd]{pmdco:characteristic} relations, based on the experiment workflow, or on both, depending on the level of filtering we want to apply.

\subsection{Modelling Experiment Plans}
\label{sec:handover_workflow_models}

As part of our implementation, we have designed an ontological representation of \emph{experiment plans}. Modelled as a \mbox{\tcbox[crc]{crc:HandoverWorkflowModel}} instance, an experiment plan is designed to indicate the \emph{ideal} sequence of projects a given research object should go through as steps, and which activities should be performed within each one. In comparison to the experiment workflows previously detailed in Section~\ref{sec:modeling_handover_workflows}, an experiment plan is used to define present and future activities without being backed by experiment results nor handover events. Thus, the steps of an experiment plans act akin to \mbox{\tcbox[crc]{crc:HandoverGroup}} entities, linking directly to \tcbox[crc]{crc:CharacterizationActivity} entities and thereby skipping the \mbox{\tcbox[crc]{crc:Handover}} layer. 

Conversely, a \tcbox[crc]{crc:HandoverWorkflowModelInstance} declares research object assignments to each of the steps of an experiment plan, thus allowing for the reuse of their definitions. This results in a highly flexible representation, supporting branching into multiple objects and steps, or introducing new objects in an intermediary step. This way, a single experiment plan can encompass multiple experiment workflows from a set of (typically related) research objects.

\begin{figure}[t]
    \centering
    
    \begin{minipage}[t]{0.394\linewidth}
         \includegraphics[width=\linewidth]{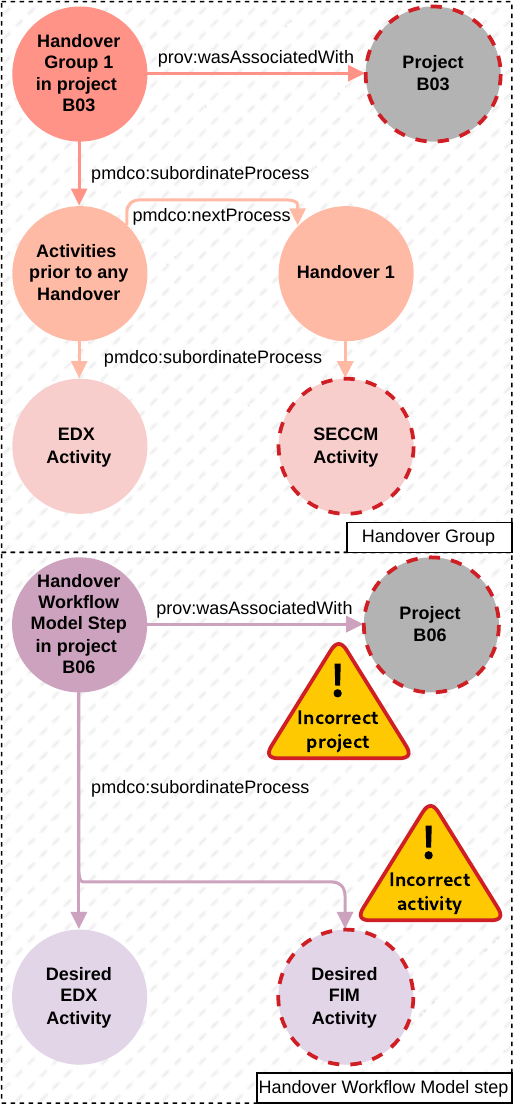}
    \end{minipage}
    \begin{minipage}[t]{0.59\linewidth}
        \begin{titlebluebox}{SHACL shape}
            \begin{minted}[fontsize=\scriptsize,escapeinside=||]{turtle}
crc:example_validation_shape
  a sh:NodeShape ;
  sh:targetNode handover_group:1 ;

  sh:property [
    sh:path prov:wasAssociatedWith ;
    sh:in (project:B06) ;
    sh:message "Incorrect project"
  ] ;
  
  sh:property [
    sh:path (pmdco:subordinateProcess 
            [sh:zeroOrMorePath pmdco:nextProcess] 
            pmdco:subordinateProcess) ;
    sh:qualifiedValueShape [
      sh:class crc:EDXProcess ;
    ] ;

    sh:qualifiedMinCount 1 ;
    sh:message "EDX has not been performed"
  ] ;
  
  sh:property [
    sh:path (pmdco:subordinateProcess 
            [sh:zeroOrMorePath pmdco:nextProcess] 
            pmdco:subordinateProcess) ;
    sh:qualifiedValueShape [
      sh:class crc:FIMProcess ;
    ] ;

    sh:qualifiedMinCount 1 ;
    sh:message "FIM has not been performed"
  ] .
\end{minted}
        \end{titlebluebox}
    \end{minipage}
    \caption{Validation example of a handover group against its corresponding experiment plan step with two mismatches, and its generated SHACL shape.}
    \label{fig:validation_example}
\end{figure}
\begin{figure}[t]
    \centering
    \begin{minipage}[c]{0.49\linewidth}              
    \frame{\includegraphics[width=\linewidth]{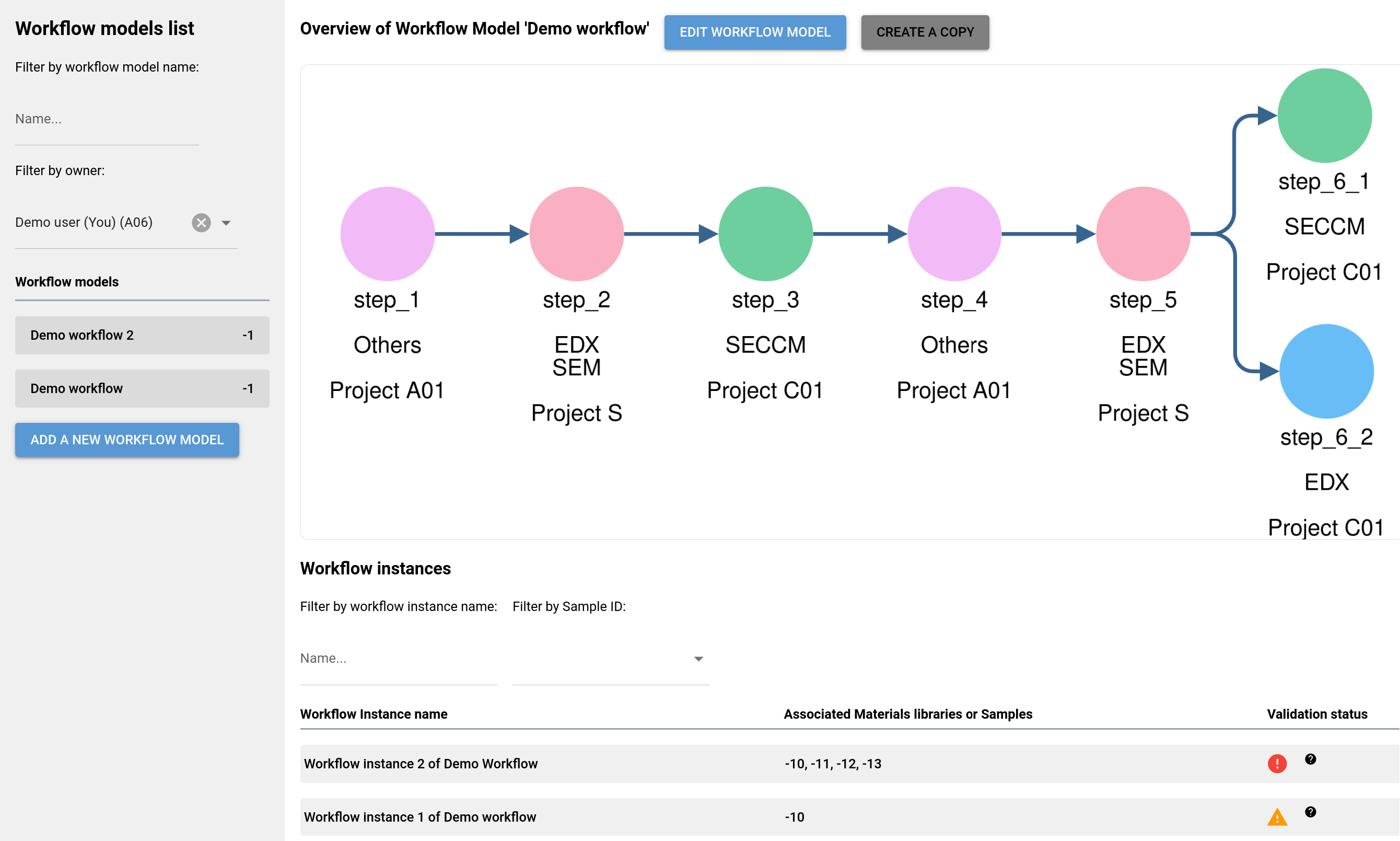}}
    \end{minipage}
    \begin{minipage}[c]{0.49\linewidth}
    \frame{\includegraphics[width=\linewidth]{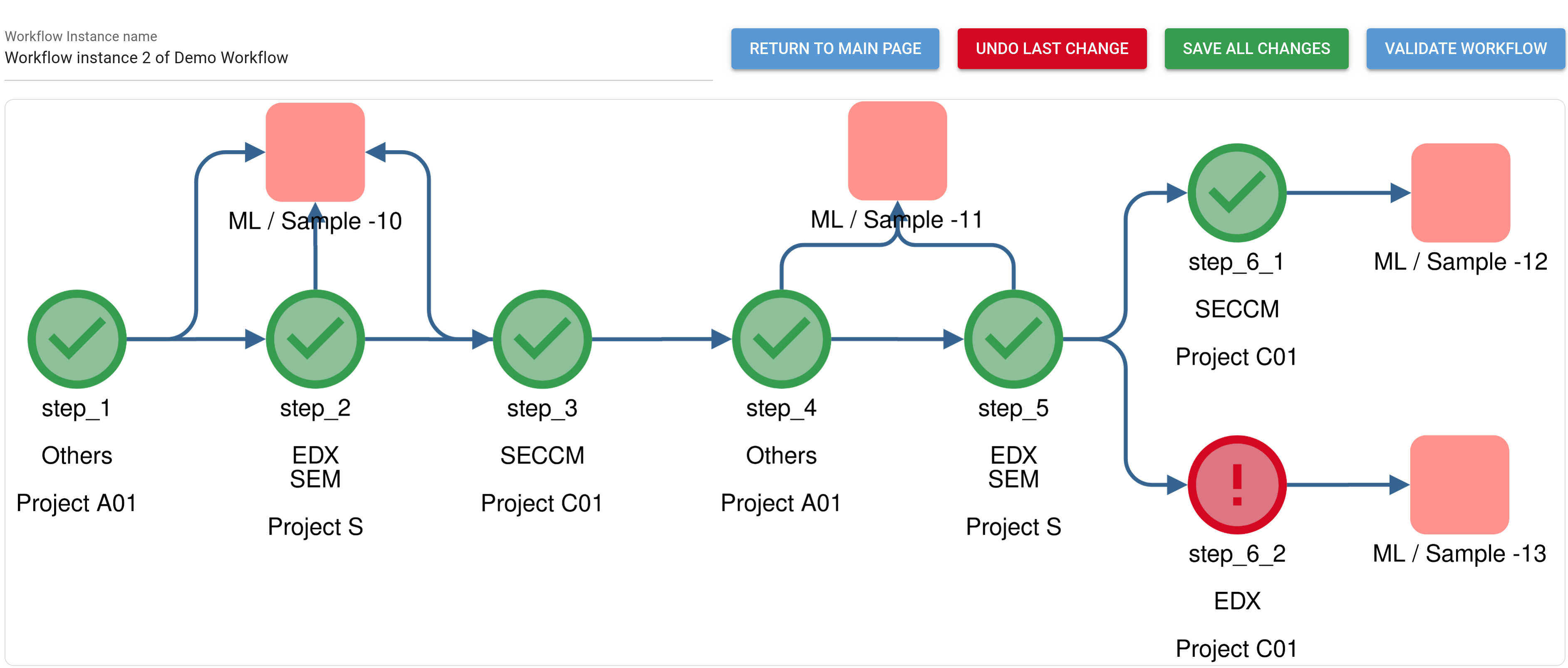}}
    \end{minipage}
    \caption{User-facing experiment plans dashboard showing a workflow model (left) and the validation results overview of one of its instances (right).}
    \label{fig:webui_screenshots}
\end{figure}

\boldparagraph{Validating Experiment Plans} Acting as ideal experiment workflows, the experiment plans serve to validate the conformance of the actual experiment workflow counterparts logged by the researchers. To achieve this, we have developed a validation API that, given an experiment plan and its corresponding instance, sequentially pairs the steps of the plan to the chain of \tcbox[crc]{crc:HandoverGroup} instances of the associated research objects\footnote{If an object is newly introduced in an intermediate step, it will be paired from its first handover group onwards. This mechanism allows associating parts of the same experiment plan to different research objects.}. For each one of these pairings, the system generates a corresponding SHACL~\cite{shacl_spec} shape to be validated via pySHACL~\cite{pySHACL}. Currently, these SHACL shapes are pre-defined and templated, with its details filled according to the configuration of the corresponding experiment plan step (e.g., the project the handover group should take place in, and the desired activities to be performed). An example of the modelling and the validation process can be found in Figure~\ref{fig:validation_example}.

This system is currently available as a web interface acting as an \textit{experiment plans dashboard}, where end users can define a plan of activities for a set of research objects they are interested in. Acting as a journal, it then tracks the status of each one of them with respect to these plans by using the validation API. An example of the dashboard can be found in Figure~\ref{fig:webui_screenshots}.

Note that our approach constitutes a novel application of validation languages in RDF. In this case, we ensure that the handover workflows logged in the system \emph{semantically} refer to a reality represented by an experiment plan, with SHACL being used to establish their equality criteria. This contrasts with typical use-cases aiming to validate the the materialized entities of the KG (e.g., ensuring there are no duplicate activities under a single handover)\footnote{The handover workflows are already structurally sound by virtue of them being rigidly materialized by the mappings. This aspect will be complemented in the future by expressing our ontology as SHACL shapes.}. Moreover, the validation is purely user-driven, with its coverage depending on the experiment plans that our users define.



\section{CCSS Knowledge Graph Construction}
\label{sec:kgc}
Given that, currently, our sole data source is MatInf's relational database, using a mapping engine is the most straightforward choice. 
These mappings are written as a collection of 29 YARRRML~\cite{yarrrml} files organized into categories (e.g., users, experiment workflows or composition data).
Since, as described in Section~\ref{sec:db_schema}, the database contains object types to be modelled in the ontology, some mappings are \emph{templated} according to preset subsets of these types. Likewise, other mappings automatically inject new (sub-)classes in the ontology, as is the case of the measurement format types. This offers us a degree of consistency of the mappings when this measurement formats list is updated: new ones will be grouped among others within a \emph{catch-all} activity for unknown classes until it is added to the mappings, but will still be present in the ontology and consequently queryable. 

These YARRRML mapping files are coalesced into one single RML~\cite{rmlspec} mapping, which is then executed via Morph-KGC~\cite{arenas2024morph}. Following this, the 3-layer experiment workflows hierarchy representation (cf.  Section~\ref{sec:modeling_handover_workflows}), is finalized in an entity-refinement step in charge of creating \tcbox[crc]{crc:HandoverGroup} instances. This reduces the complexity of the mappings and improves their maintainability, as formulating such transformations in SPARQL is trivial compared to including them within the SQL queries\footnote{Currently, materializing the experiment workflows without handover groups requires running 10 SQL queries, with 28 JOIN statements and 13 subqueries in total.}. 
The entity-refinement step also performs additional ontology integrations dynamically, currently transforming chemical element string literals (e.g. ``Au'', or ``Pt'') into ChEBI~\cite{chebi} entities via SPARQL queries, and thus avoiding writing hardcoded \emph{string-to-IRI} correspondences. 

The benefits of this entity-refinement step are the reason why we opted for an ETL approach instead of OBDA~\cite{ontop}. Moreover, this allows us to query the data natively in SPARQL, instead of through query rewriting. However, this requires running this pipeline periodically to keep the KG up to date with the MatInf instance. To maintain data consistency across executions, all entity IRIs in the KG are minted based on the unique, permanent object identifiers in MatInf. 

Finally, a collection of validation tests ensures the correctness of the mapping process against a collection of MatInf instances, by comparing the outputs to manually created KG counterparts.  A summary of ontology and KG statistics for the current deployment can be found in Table~\ref{tab:stats}.

\begin{table}[t]
\centering
\scriptsize
\caption{CCSS ontology and KG statistics (April 2026).}
\vspace{2mm}
\centering
\begin{minipage}{0.39\textwidth}
\centering
    \begin{tabular}{@{}lr@{}}
    \toprule
    \multicolumn{2}{c}{Ontology statistics} \\ \midrule
    New (sub-)classes                          & 50        \\
    Equivalent classes                         & 6         \\
    New properties                             & 23        \\
    Equivalent properties                      & 14        \\ \bottomrule
    \end{tabular}
\end{minipage}%
\hfill
\begin{minipage}{0.59\textwidth}
\centering
    \begin{tabular}{@{}lr@{}}
    \toprule
    \multicolumn{2}{c}{KG statistics} \\ \midrule
    RDF triples                         & 1,177,153 \\
    Research object entities            & 776       \\
    Research object handover entities   & 1,854     \\
    Experiment result entities          & 4,916     \\ \bottomrule
    \end{tabular}
\end{minipage}
\label{tab:stats}
\end{table}

\section{Evaluation}
\label{sec:evaluation}
As part of the analysis of the ETL process and the resulting KG itself, we have designed three evaluations,  
whose setup is detailed in \S\ref{sec:eval_setup}. 
First, we have designed a performance evaluation of the mappings execution (\S\ref{sec:mat_times}). 
Second, we have compared the performance of equivalent SQL and SPARQL query pairs on their respective data sources and discussed their complexity (\S\ref{sec:query_times}). 
Lastly, we discuss the impact and uptake of our solution in the CRC 1625 (\S\ref{sec:uptake_impact}), and summarize the lessons learned  (\S\ref{sec:lessons_learned}).

\subsection{Evaluation Setup}
\label{sec:eval_setup}
To evaluate the performance of our ETL approach, we have developed a synthetic data generator that creates valid MatInf database instances of different sizes,  allowing us to stress-test our approach as the underlying database grows. 
Given an initial number of materials libraries to instantiate, this generator leverages MatInf's categorization of every object into a type to simulate the statistical distribution of related objects and their types.
This includes their splitting into samples, randomized handover chains for each of them, provenance metadata and the creation of complex EDX compositions. Their individual probabilities are then scaled by a value $k$ to simulate an increased user activity on MatInf, using a non-saturating soft scaling function ${P_{scaled} = P^{(\frac{1}{k})}}$. 

All tests were run on an Ubuntu 24.04 server, using an AMD EPYC 9224 CPU (24c / 48t), 566 GiB RAM and a 7 TiB SATA SSD. Microsoft SQL Server 2022 and OpenLink Virtuoso OS Edition 7.2 were used as the SQL and RDF stores, respectively.


\subsection{Knowledge Graph Construction Performance}
\label{sec:mat_times}
\begin{figure}[t]
    \centering
    \begin{subfigure}{0.98\linewidth}
        \centering
        \includegraphics[width=\linewidth]{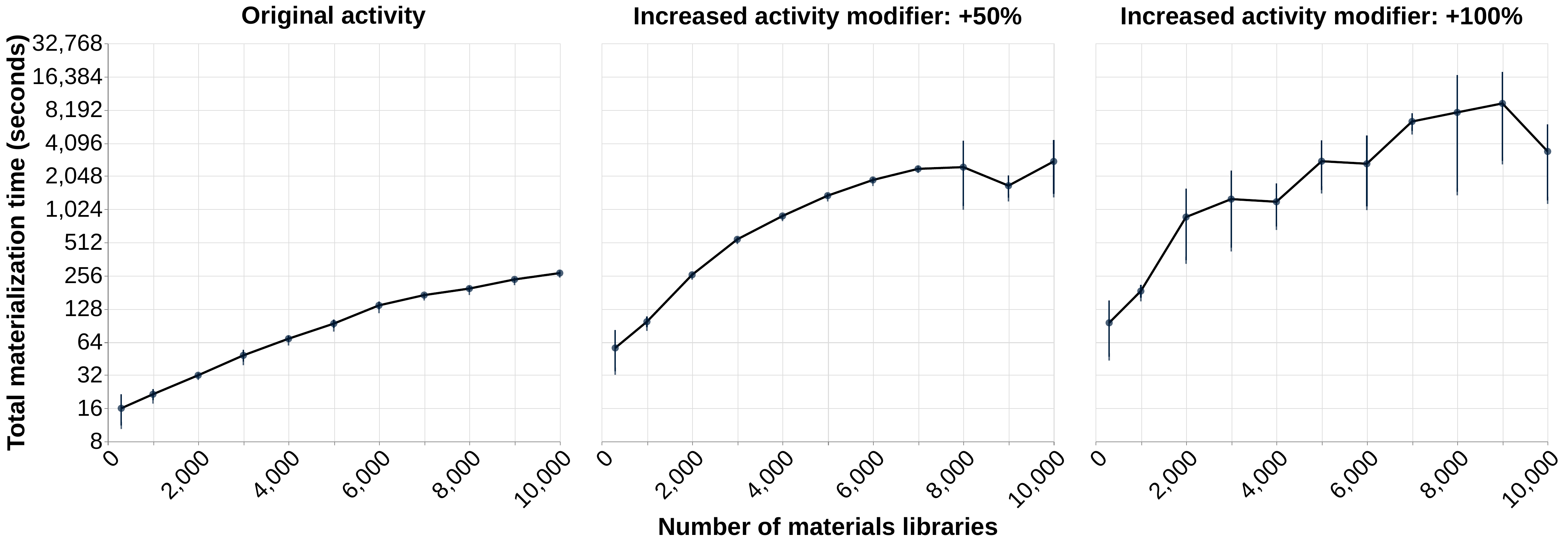}
        \caption{Materialization time per number of materials libraries.}
        \label{fig:mat_times_horizontal_1}
    \end{subfigure}
    \begin{subfigure}{0.98\linewidth}
        \centering
        \includegraphics[width=\linewidth]{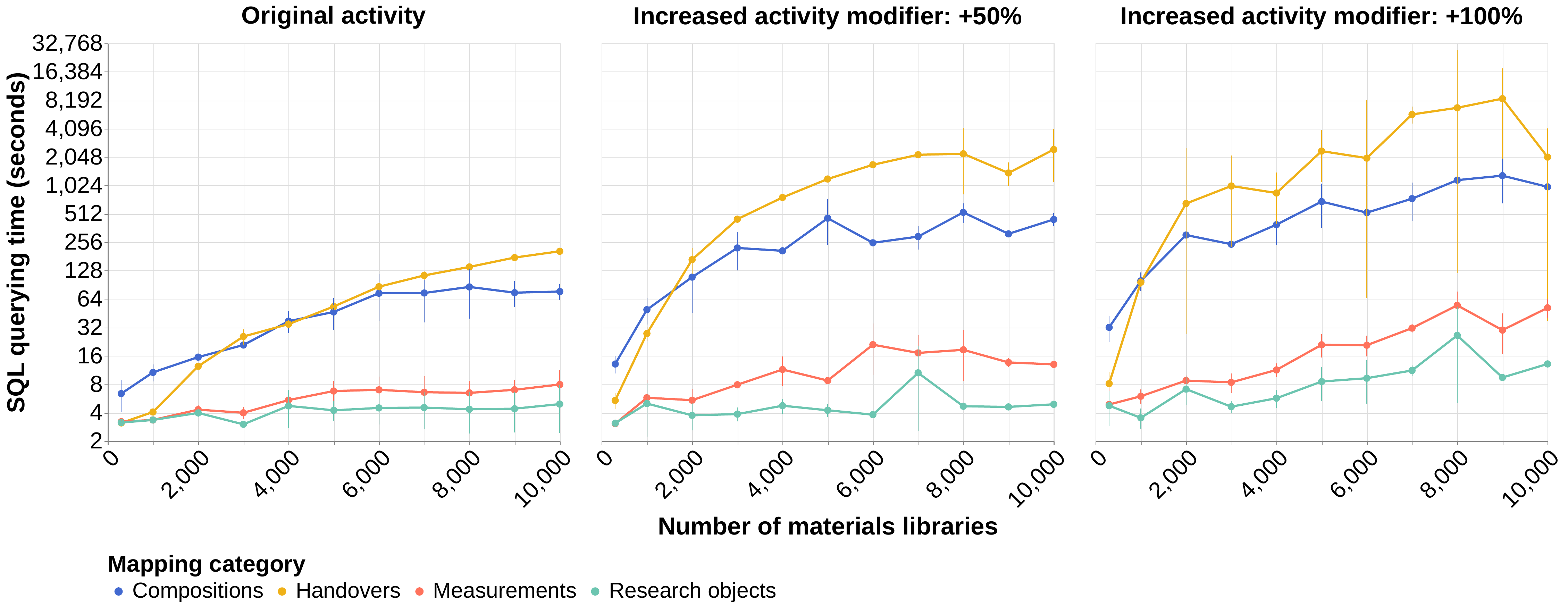}
        \caption{Mappings SQL querying time per number of materials libraries and data category.}
        \label{fig:mat_times_horizontal_2}
    \end{subfigure}
    
    \caption{KG construction runtime (log scale) under our simulated activities.} 
    \label{fig:mat_times_horizontal}
\end{figure}

We used the synthetic data generator to evaluate the mappings execution (SQL query execution and RDF conversion) and entity-refinement times from 299 starting materials libraries, corresponding to their total count as of February 2026, up to $10,000$ in steps of $1,000$. This has been performed for the original distribution of data $(k=1.0)$ and for two increased activity modifiers, $k=1.5\ (+50\%)$ and  $k=2.0\ (+100\%)$, with 3 repetitions for each test. As found in Figure~\ref{fig:mat_times_horizontal_1}, we can observe that materialization times are highly stable except for the largest numbers of starting materials libraries and highest activity levels. This is explained by instead analysing the query execution times in Figure~\ref{fig:mat_times_horizontal_2} across the different research data categories: the runtime of the mappings themselves is negligible against that of the SQL queries, due to their complexity and volume of retrieved data. The entity-refinement step to consolidate the experiment workflow representation and ChEBI's integration took an average of $43.52s$ ($k=1.0$), $163.61s$ ($k=1.5$), and $373.08s$ ($k=2.0$). The KG size scaled linearly, with $10.02\ \text{M}$, $47.10\ \text{M}$, and $116.88\ \text{M}$ triples on average across all tests for the same $k$ values. 

Conversely, in our production environment, consisting of a weaker virtual machine communicating with the database server remotely, executing the entire KG construction pipeline only requires 40 seconds. At the same time, our low volume of transactions does not require a strict consistency window; nevertheless, we re-materialize the KG every 15 minutes. Since we don't expect to reach the higher activity scenarios, these results are satisfactory.

\subsection{Query Benchmarking and Complexity}
\label{sec:query_times}
\begin{figure}[t]
    \centering
    \includegraphics[width=0.96\linewidth]{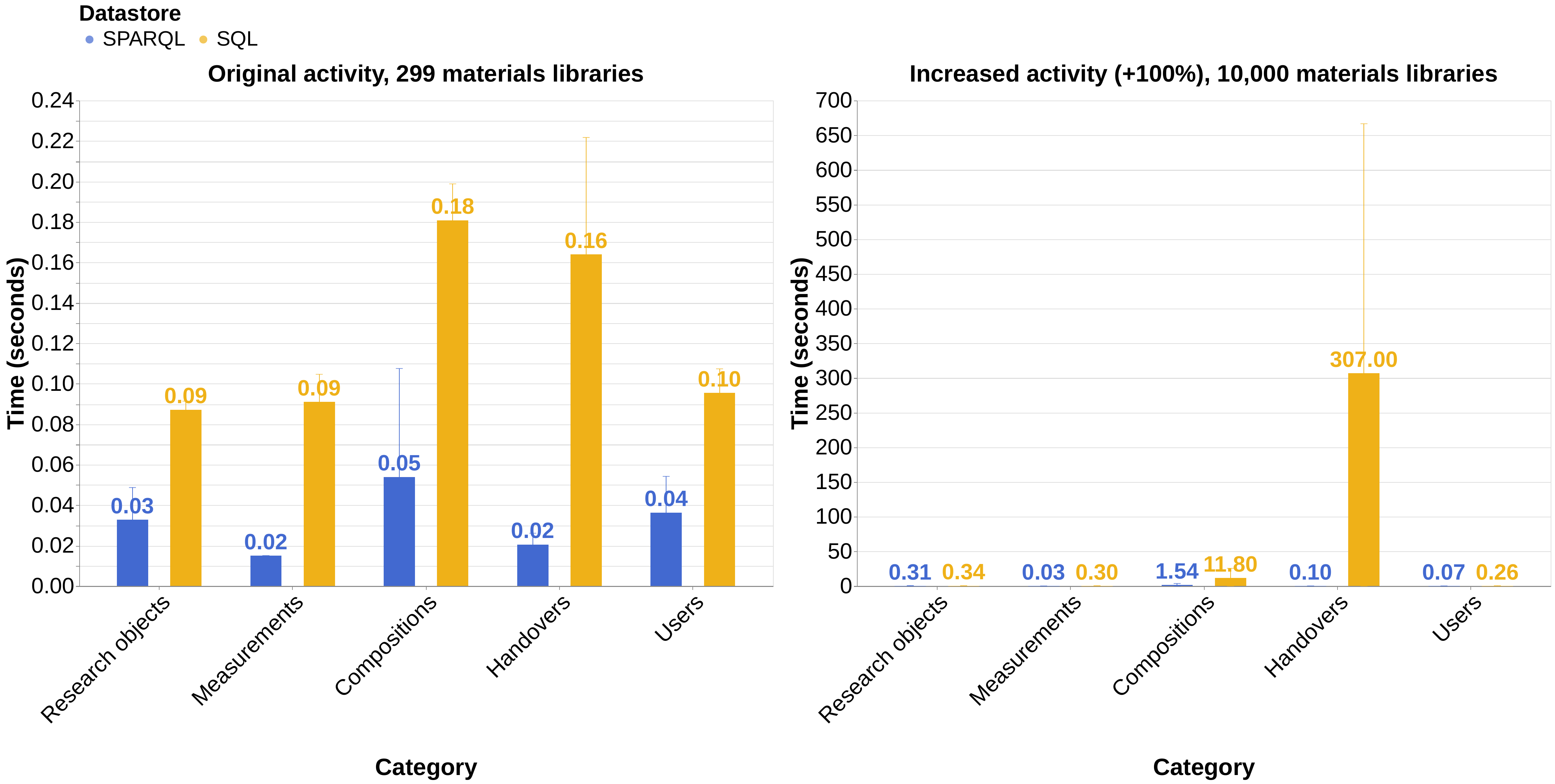}
    \caption{Querying benchmark results for the smallest and largest databases. Execution times of queries belonging to the same category are averaged.}
    \label{fig:querying_benchmark}
\end{figure}

To ensure that the KG can sustain complex data access needs, we have designed a benchmark consisting of equivalent SQL and SPARQL queries across the main categories of data\footnote{Every SPARQL-SQL query pair yields the same records. If a SPARQL query yields an IRI, its SQL equivalent will yield the internal ID of the corresponding object. We additionally removed the network latency from the execution times.}: research objects, measurements, compositions, experiment workflows and, additionally, users and projects. As we can observe for the smallest and largest synthetic databases in Figure~\ref{fig:querying_benchmark}, the querying performance is on par between the two data sources, with the KG showing an advantage on queries based on experiment workflows. This is due to the natural modelling of this data category as a graph: While SPARQL excels at this, the SQL queries require self-joins and nested sub-queries, which are expensive to execute.

\label{sec:query_legibility}
With the aim of lowering the barrier of entry to SPARQL querying, we additionally maintain a parallel namespace for the reused PMDco properties and classes via OWL equivalence axioms. This way, end users can query the graph by exclusively using the same vocabulary from our user-facing tools and MatInf (e.g., \tcbox[crc]{crc:nextStep} is equivalent to \tcbox[pmd]{pmdco:nextProcess}).

\subsection{Uptake and Impact}
\label{sec:uptake_impact}
Our implementation represents the first production-ready iteration of a semantic layer that complements MatInf, joining the ongoing efforts towards the inclusion of these technologies in the RDM ecosystem as a whole. It has been currently rolled out for the 73 members and collaborators of the CRC 1625 registered in MatInf, of which there are 50 active data contributors. 

As a Collaborative Research Center (CRC), however, the development and rollout of our semantic layer's features differs from other scenarios. As a project set to be evaluated and extended throughout three 4-year funding periods, our aim is to establish services in the medium- to long-term. Consequently, in the current initial phase, our focus is on maintaining a constant dialogue with our users to closely uncover and develop previously unexplored use cases. Due to this, our technology value is not currently driven by concrete metrics, but rather by how our users are willing to interact and exploit research data in the future. As a result, our assessment mechanisms are our user's interest, their feedback about our proposals, and their feature requests.

\boldparagraph{SPARQL-based Data Querying}
The KG, currently deployed internally for our users, provides direct querying access through web UIs and APIs. By providing demos and courses, we have shown our researchers that interacting with their data via SPARQL queries is not only simpler but \emph{possible} for them, as the previous alternative was SQL-based access. Currently, we plan on replacing or complementing the current SQL-based access scripts with SPARQL alternatives. This will become crucial as more data is generated, as the different projects will jointly query and analyze their experimental data.


\boldparagraph{Experiment Plans}
Our feedback rounds with CRC 1625 MatInf users have highlighted the usefulness of the experiment plans. As a result, we have already obtained feature requests for grouping these plans on a per-publication basis. This will additionally enable us to export them alongside the experimental \mbox{(meta-)data} as, e.g., RO-Crates~\cite{soiland-reyesPackagingResearchArtefacts2022}. Its final deployment, taking all feature requests into account, will then be used throughout the final experimental work of this current phase, and further maintained during the next phases.

\subsection{Lessons Learned}
\label{sec:lessons_learned}
One of our main development goals is to go beyond a basic queryable KG, data publication or reusability. Instead, we aim to release tools that are 1)~specifically tailored for our users in terms of functionalities, vocabulary and complexity, and 2)~beneficial for them. This has been shown to be only possible through constant dialogue, demos, courses and feedback sessions, which allow us to refine and repurpose our use-case proposals.

These tools, acting as \emph{visible} artifacts of our work within RDM, may eventually foster better logging of all data in MatInf as a side effect, as they allow present and future researchers to directly interact with the collaborative history of experimental data and workflows logged in MatInf.


On the technical side, we wish to maintain our ETL approach, due to the simpler maintainability of the mappings and lack of possible query rewriting costs that an ODBA approach may carry. If materialization times increase in the future, we aim to investigate the inclusion of data stream mappings of new and modified objects in MatInf, or adopting incremental KG construction approaches~\cite{van2026incremental}.

\section{Related Work}
\label{sec:rel_work}

\noindent
\textbf{Semantic Web Technologies in RDM Systems}
To the best of our knowledge, there are no production-ready RDM systems that are either purely RDF-based or use a tightly integrated semantic layer over both the data and metadata, as in our case. Nonetheless, we can find several approaches integrating these technologies in RDM systems. 

In this line, Coscine~\cite{coscine} represents metadata in RDF and validates it with SHACL. Similarly, openBIS provides an integration with semantic web technologies by allowing to semantically annotate types and properties within the system\footnote{\href{https://openbis.readthedocs.io/en/20.10.12-plus/software-developer-documentation/apis/java-javascript-v3-api.html\#semantic-annotations}{Semantic annotations in openBIS} (Last accessed: July 2026).}. As detailed by Kockmann et al.~\cite{nfdi4catwhitepaper}, LARAsuite\footnote{\href{https://gitlab.com/LARAsuite}{LARAsuite} (Last accessed: July 2026).} offers a semantic layer for metadata, while Adacta\footnote{\href{https://github.com/adacta-rdm/adacta}{Adacta} (Last accessed: July 2026).} links Electronic Lab Notebooks (ELNs) and other resources via common ontologies, and NOMAD\footnote{\href{https://github.com/FAIRmat-NFDI/nomad}{NOMAD} (Last accessed: July 2026).} allows assigning ontology terms to experimental data schemas.

\smallbreak
\noindent
\textbf{Semantic Experiment Workflows}
With a focus on computational workflows, we can find the works by Garijo et al.~\cite{garijoNewApproachPublishing2011} introducing abstract and executable computational workflows, and the use of ontologies for modelling workflow provenance metadata with OPMW-PROV~\cite{opmw-prov}, whose core ideas are currently implemented in current TLOs and MLOs. More recently, Dias et al.~\cite{diasMAESTROLightweightOntologybased2024} use ontologies to represent experiment workflows and derive execution plans via reasoning, and Franz et al.~\cite{rotersStahlDigitalOntologyBasedWorkflows2025} express computational workflows from pyiron\footnote{\href{https://pyiron.org/}{pyiron} (Last accessed: July 2026).} via ontologies, making them interoperable with an existing, semantically enriched data collection.

Due to the mixed computational and experimental nature of our work, we use a representation of experiment workflows that slightly deviates from traditional computational workflow definitions in the literature. To the best of our knowledge, there are also no works focused on validating RDF-based, non-computational experiment plans. 
Nonetheless, Miksa et al.~\cite{miksaUsingOntologiesVerification2017} use ontologies to express validation requirements of computational workflows, and SHACL is also used in RDM to validate experimental data~\cite{ozcepModelingAccessingSmart2025} or for data ingestion forms~\cite{kirchnerOntologybasedDescriptionNano2025}.  

\section{Conclusion}
\label{sec:conclusion_future_work}
In this work, we showcase the work performed as part of the 
Collaborative Research Center (CRC) 1625 to create a tightly integrated semantic layer over MatInf, a domain-specific Research Data Management (RDM). Aside from the flexible data access benefits brought by this integration, we showcase how the graph nature of research data and experiment workflows unveils use cases that would have been otherwise complex to implement in a relational data source. 

This is demonstrated with the experiment plans, which allow end users to plan, reuse, share and track the progress of their day-to-day activities. Despite developing these solutions in the context of the CRC 1625, we aim to formalize and apply our current experiment plans validation strategy to general RDF workflows. Moreover, we plan to expand the CCSS ontology to other projects using MatInf, and to align it with additional ontologies. Currently, we are investigating an update to the newer PMDco v3.0.x, and additionally expressing our experiment workflows with PKO~\cite{PKO}. 

\boldparagraph{Supplemental Material}
The ontology, source code, documentation and evaluation results are available \href{https://github.com/DE-TUM/CRC1625}{on GitHub}. Additional figures (e.g., resource consumption logs or fine-grained results) are also available. A live version of the deployment, with demo user access, is accessible at \mbox{\href{https://kg.crc1625.mdi.ruhr-uni-bochum.de/}{kg.crc1625.mdi.ruhr-uni-bochum.de}}. Finally, the CCSS ontology is also available at \mbox{\href{https://purl.org/ccss-ontology}{https://purl.org/ccss-ontology}}.

\boldparagraph{Acknowledgments}
This work has been funded by the Deutsche Forschungsgemeinschaft (DFG, German Research Foundation) - SFB 1625 - 506711657, subprojects A06 and INF. 

\FloatBarrier
\clearpage
\section*{Declaration of use of Generative AI}
During the writing of this work, generative LLMs (ChatGPT and Gemini) have been exclusively used to perform small adjustments to the text (e.g., correcting grammatical mistakes, suggesting synonyms, or improving English idioms). While programming the system, they have been likewise used exclusively to suggest improvements to the code or solve errors. All ideas, text, figures, images, and code are of original authorship.

\bibliographystyle{splncs04}
\bibliography{biblio}

\end{document}